\documentclass[aps,prx,twocolumn,showpacs]{revtex4-1}
\usepackage{amsfonts}
\usepackage{amsmath}
\usepackage{amsthm}
\usepackage{graphics}
\usepackage{graphicx}
\usepackage{subfigure}
\usepackage{amssymb}
\usepackage{graphics}
\usepackage{graphicx}
\usepackage{epstopdf}
\usepackage{color}
\usepackage{subfigure}
\usepackage{pgfplots,tikz}
\usepackage{url}
\usepackage{float}
\usepackage{soul}

\usepackage{verbatim}
\def\x{{\bf x}}

\def\q{\textbf{q}}

\newcommand\dertt[1]{ \frac{\partial{ #1}}{\partial t} }

\def\Vp{V_{\rm p}}
\def\Mp{M_{\rm p}}
\def\Rp{a_{\rm p}}

\begin{document}

\title{How trapped particles interact with and sample superfluid vortex excitations}
\author{Umberto Giuriato}
\affiliation{
Universit\'e C\^ote d'Azur, Observatoire de la C\^ote d'Azur, CNRS, Laboratoire Lagrange, Nice, France}
\author{Giorgio Krstulovic}
\affiliation{
Universit\'e C\^ote d'Azur, Observatoire de la C\^ote d'Azur, CNRS, Laboratoire Lagrange, Nice, France}
\author{Sergey Nazarenko}
\affiliation{Universit\'e C\^ote d'Azur, Institut de Physique de Nice, CNRS, Nice, France}


\begin{abstract}
Particles have been used for more than a decade to visualize and study the dynamics of quantum vortices in superfluid helium. In this work we study how the dynamics of a collection of particles set inside a vortex reflects the motion of the vortex. We use a self-consistent model based on the Gross-Pitaevskii equation coupled with classical particle dynamics. 
We find that each particle oscillates with a natural frequency proportional to the number of vortices attached to it. We then study the dynamics of an array of particles trapped in a quantum vortex and use particle trajectories to measure the frequency spectrum of the vortex excitations.  Surprisingly, due to the discreetness of the array, the vortex excitations measured by the particles exhibits bands, gaps and Brillouin zones, analogous to the ones of electrons moving in crystals. We then establish a mathematical analogy where vortex excitations play the role of electrons and particles that of the potential barriers constituting the crystal. We find that the height of the effective potential barriers is proportional to the particle mass and the frequency of the incoming waves. We conclude that large-scale vortex excitations could be in principle directly measured by particles and novel physics could emerge from particle-vortex interaction.
\end{abstract}
\maketitle

\section{Introduction}
When a fluid composed of bosons is cooled down, a spectacular phase transition takes place. The system becomes superfluid and exhibits exotic physical properties. Unlike any classical fluid, a superfluid flows with no viscosity. This is an intriguing example of the manifestation of pure quantum mechanical effects on a macroscopic level. The first discovered superfluid is liquid helium $^4\mathrm{He}$ in its so-called phase $\mathrm{II}$, below the critical temperature $T_\lambda\simeq 2.17\,\textrm{K}$. In one of the first attempts of describing the behavior of superfluid helium, London suggested that superfluidity is intimately linked to the phenomenon of Bose-Einstein condensation (BEC) \cite{London}. In the same years, Landau and Tisza independently put forward a phenomenological two-fluid model, wherein superfluid helium can be regarded as a physically inseparable mixture of two components: a normal viscous component that carries the entire entropy and an inviscid component with zero entropy \cite{Landau,Tisza}.

Because of its intrinsic long-range order, a superfluid can be described by a macroscopic complex wave function. A stunning quantum-mechanical constraint is that vortices appear as topological defects of such order parameter. In three dimensions, such defects are unidimensional structures, usually referred to as quantum vortices. Indeed, the circulation (contour integral) of the flow around a vortex must be a multiple of the Feynman-Onsager quantum of circulation $h/m$, where $h$ is the Planck constant and $m$ is the mass of the Bosons constituting the fluid \cite{Feynman}. Such peculiarity is necessary to ensure the monodromy of the wave function. In superfluid helium, quantum vortices have a core size of the order of an Angstrom. At low temperatures, below 1\,K, the normal component is negligible and vortices are stable and do not decay by any diffusion process, unlike their classical counterparts. The understanding of superfluid vortex dynamics has a direct impact on many interesting, complex non-equilibrium multi-scale phenomena, such as turbulence \cite{vinen2002quantum,Paoletti_QT,BarenghiIntroQunatumTurbu}.

Most of the experimental knowledge on superfluid vortices is based on indirect measurement of their properties. The early efforts in the observation of quantized vortices were made in the framework of rotating superfluid helium, by using electron bubbles (ions) as probes \cite{donnelly1991quantized}. Since then, impurities have been extensively used to unveil the dynamics of superfluid vortices. An important breakthrough occurred in 2006, when micrometer sized hydrogen ice particles were used to directly visualize superfluid helium vortices \cite{bewley2006superfluid}.  Thanks to pressure gradients, particles get trapped inside quantum vortices and are subsequently carried by them. Hence, it has been possible to observe vortex reconnections and Kelvin waves (helicoidal displacements that propagate along the vortex line) by means of standard particle-tracking techniques \cite{FondaKWExp}. Furthermore, the particle dynamics unveiled important differences between velocity statistics of quantum and classical turbulent states \cite{PaolettiVelStat2008,LaMantiaPRBVelStat2014}. In experiments, such particles are used as tracers, despite their very large size compared to the vortex core. Therefore, it is of the utmost importance that the mechanisms driving their dynamics are fully comprehended. Specifically, how well is vortex dynamics reflected by the motion of the particles trapped in it? How much do their presence in the core modify the propagation of Kelvin waves? Would they affect the reconnection rates?
 
Describing the interaction of particles with isolated vortex lines or complex quantum vortex tangles is not an easy task. Depending on the scale of interest, there are different theoretical and numerical models that can be adopted.
A big effort has been made in adapting the standard dynamics of particles in classical fluids to the case of superfluids described by two-fluid models \cite{ParticlesNewcastle,ReviewNewcastle}. This is a macroscopic model in which vorticity is a coarse-grained field and therefore there is no notion of quantized vortices. A medium-scale description is given by the vortex filament model, where the superfluid is modeled as a collection of lines that evolve following Biot-Savart integrals. In this approximation, circulation of vortices is by construction quantized but reconnections are absent and have to be implemented via some ad-hoc mechanism. Finite size particles can be studied in the vortex filament framework but the resulting equations are numerically costly and limited \cite{CloseParticleNewcastle}. A microscopic approach consists in describing each impurity by a classical field in the framework of the Gross-Pitaevskii model \cite{PitaevskiiTheory,RicaRoberts,VilloisBubble}. In principle, such method is valid for weakly interacting BECs, and is numerically and theoretically difficult to handle if one wants to consider more than just a few particles. In the same context, an alternative possibility is to assume classical degrees of freedom for the particles, while the superfluid is still a complex field obeying the Gross-Pitaevskii equation. This idea of modelling particles as simple classical hard spheres has been shown to be both numerically and analytically very powerful \cite{ActiveWiniecki,ShuklaParticlesPRA2016,giuriato2018clustering,GiuriatoApproach}. In particular, such minimal and self-consistent model allows for simulating a relatively large number of particles, and describes well the particle-vortex interaction \cite{GiuriatoApproach}. Although formally valid for weakly interacting BECs, it is expected to give a good qualitative description of superfluid helium.

In this paper we investigate how particles trapped in quantum vortices interact with vortex excitations and in particular how well they can be used to infer properties of superfluid vortices. We use the Gross-Pitaevskii equation coupled with inertial and active particles obeying classical dynamics to answer this question. We first address how the Magnus force acting on trapped particles induces oscillations at a certain natural frequency. This quantity may be experimentally measured to determine the number of vortices composing a polarized bundle (see a discussion later in this paper). Secondly, in order to understand the effect of particle inertia, we analyze the spectrum of vortex excitations in the case when a continuous distribution of mass is contained inside the vortex core. 
Then, we study an array of particles trapped inside a vortex, in a setting similar to the one observed in experiments. Surprisingly, the dispersion relation of vortex waves measured by the particles is found to contain band gaps and the periodicity typically observed in the energy spectra of solids. We explain the numerical observation applying the concepts used in the standard Kronig-Penney model \cite{KronigPenney,kittel1976introduction}, that describes the motion of electrons in a unidimensional crystal. Finally, based on our results, we discuss in which regimes particles could be reliably used to sample vortex excitations.

\section{Theoretical Background}
\subsection{Model for superfluid vortices and active particles}

We consider a superfluid at very low temperature containing $N_\mathrm{p}$ spherical particles of mass $\Mp$ and radius $a_\mathrm{p}$. We describe the system by a self-consistent model based on the three-dimensional Gross-Pitaevskii equation. The particles are modeled by strong localized potentials $\Vp$, that completely deplete the superfluid up to a distance $a_\mathrm{p}$ from their center position $\mathbf{q}_i$. Particles have inertia and obey a Newtonian dynamics. 
The Hamiltonian of the system is
\begin{eqnarray}
H&=&\int\left(\frac{\hbar^2}{2m}|\nabla\psi|^2+\frac{g}{2}|\psi|^4 + {\sum_{i=1}^{N_\mathrm{p}}V_\mathrm{p}(\mathbf{r}-\mathbf{q}_i)|\psi|^2} \right)\mathrm{d}\mathbf{r}\nonumber\\
&+& {\sum_{i=1}^{N_\mathrm{p}}\frac{\mathbf{p}_i^2}{2M_\mathrm{p}}} 
+ {\sum_{i<j}^{N_\mathrm{p}}V_\mathrm{rep}^{ij}}.
\label{Eq:Hamiltonian}
\end{eqnarray}
where $\psi$ is the wave function that describes the superfluid and $m$ is the mass of the condensed bosons interacting with a $s$-wave scattering length $a_{\rm s}$, so that the coupling constant is $g=4 \pi  a_\mathrm{s} \hbar^2 /m$.  The potential  $V_\mathrm{rep}^{ij}=\varepsilon(r_0/|{\bf q}_i-{\bf q}_j|)^{12}$ is a repulsive potential of radius $r_0$ between particles. See references \cite{GiuriatoApproach, VishunathParticle1} and the next section for further details about the model.
The equations of motion for the superfluid field $\psi$ and the particle positions $\q_i=(q_{i,x},q_{i,y},q_{i,z})$ are:
\begin{eqnarray}
i\hbar\dertt{\psi} = - \frac{\hbar^2}{2m}\nabla^2 \psi + \left( g|\psi|^2-\mu\right)\psi+\sum_{i=1}^{N_\mathrm{p}}\Vp(| \x -{\bf q}_i |)\psi \label{Eq:GPEParticles}, \nonumber \\
\\
\Mp\ddot{\bf q}_i = - \int  \Vp(| \x -{\bf q}_i|) \nabla|\psi|^2\, \mathrm{d} \x+\sum_{j\neq i}^{N_\mathrm{p}}\frac{\partial}{\partial{\bf q}_i }V_\mathrm{rep}^{ij} \nonumber \\, 
\label{Eq:GP}
\end{eqnarray}
This model has been successfully used to study vortex nucleation \cite{ActiveWiniecki} and trapping of particles by quantum vortices \cite{GiuriatoApproach}. We denote by GP the Gross-Pitaevskii model without particles and by GP-P the full coupled system (\ref{Eq:GPEParticles}-\ref{Eq:GP}). 

In absence of particles, the chemical potential $\mu$ fixes the value of condensate ground state $\psi_\infty=\sqrt{{\rho_\infty}/{m}}=\sqrt{\mu/g}$.  Linearizing around this value, wave excitations are described by the Bogoliubov dispersion relation
\begin{equation}
\Omega_\mathrm{B}(k)=  c|\mathbf{k}|\sqrt{1+\frac{\xi^2 |\mathbf{k}|^2 }{2}},
\label{Eq:Bogoliubov}
\end{equation}
where $\mathbf{k}$ is the wavenumber of the excitation. Large wavelength excitations propagate with the phonon (sound) velocity $c=\sqrt{g\rho_\infty/m^2}$, while at length scales smaller than the healing length $\xi=\sqrt{\hbar^2/2g\rho_\infty }$ excitations behave as free particles.

The close relation between the GP model and hydrodynamics comes from the Madelung transformation $\psi(\x)=\sqrt{{\rho(\x)}/{m}}\,e^{i\frac{m}{\hbar}\phi(\x)}$, that maps the GP \eqref{Eq:GPEParticles} into the continuity and Bernoulli equations of a superfluid of density $\rho$ and velocity $\mathbf{v}_\mathrm{s}=\nabla\phi$. Although the superfluid velocity is potential, the phase is not defined at the nodal lines of $\psi(\x)$ and thus vortices may appear as topological defects. The simplest case corresponds to a straight quantum vortex given by
\begin{equation}
\psi_{\rm v}(x,y,z)=\sqrt{\rho_{\rm v}(x,y)/m}\,e^{i \frac{m}{\hbar}\phi_{\rm v}(x,y)}, \label{EQ:quantumVortex}
\end{equation}
where $\rho_{\rm v}(x,y)$ vanishes at the vortex core line $(0,0,z)$. The core size of a vortex is of the order of the healing length $\xi$ and the phase $\phi_{\rm v}=\frac{n_\mathrm{v}\hbar}{m}\varphi$, with $\varphi$ the angle in the $(x,y)$ plane, ensures the monodromy of the solution (\ref{EQ:quantumVortex}) only if $n_\mathrm{v}$ is an integer number.
The corresponding velocity field is $\mathbf{v}_{\rm v}=\frac{n_\mathrm{v}\hbar}{m}\frac{\hat{\varphi}}{|\mathbf{x}_\perp|}$, where $\hat{\varphi}$ is the azimuthal unit vector and $\mathbf{x_\perp}=(x,y,0)$. The circulation along a closed path $\mathcal{C}$ surrounding the vortex is therefore quantized:
\begin{equation}
\Gamma=\oint_{\mathcal{C}} \mathbf{v}_{\rm v}\cdot \mathrm{d}{\bf \mathbf{l}}=n_\mathrm{v}\frac{h}{m}= 2\pi n_\mathrm{v} \sqrt{2} c\xi.\label{Eq:VortexVelCirculation}
\end{equation}
Actually, for $|n_\mathrm{v}|>1$ vortices are structurally unstable and split into single-charged vortices. We shall consider only $n_{\mathrm v}=\pm1$ vortices.
 Note that the Bogoliubov spectrum (3) obtained in the GP framework describes well the excitations of atomic BECs, but does not match the one observed in superfluid helium. In particular, the dispersion relation never changes convexity and the roton minimum is absent. Nevertheless, the hydrodynamic description of vortices and of their large scale excitations (summarized in the following section) is similar both in helium and in the GP model.

\subsection{Frequency spectrum of superfluid vortex excitations}

\begin{figure}[h!]
\centering
\includegraphics[width=.99\linewidth]{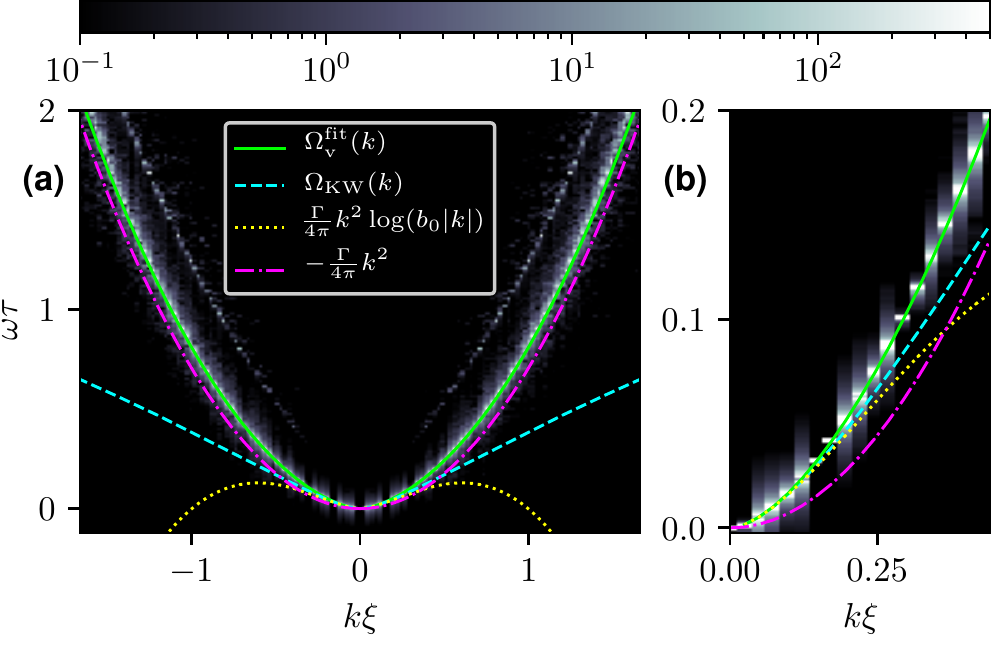}
\caption{\textbf{(a)} Spatiotemporal spectrum of a GP bare vortex loaded with small amplitude Kelvin waves. Solid green line is the fit (\ref{Eq:fit}). Dashed cyan line is KW dispersion relation (\ref{Eq:KW}). Dotted yellow line is the small-$k$ asymptotic (\ref{Eq:KWlim}), with $b_0=a_0e^{\gamma_\mathrm{E}}/2$. Magenta dash-dotted line is tha large $k$ asymptotic (\ref{Eq:freeParticles}). The resolution of the simulation is $N_\perp=N_\parallel=256$ in a computational domain of size $L_\perp=L_\parallel = 256\xi$. \textbf{(b)} A zoom close to small wave numbers.}
\label{Fig:fitcomp}
\end{figure}

Excitations are present in quantum vortices because of thermal, quantum or turbulent fluctuations. They are waves propagating along the vortex line with a certain frequency $\Omega_\mathrm{v}(k)$, where $k$ is the (one dimensional) wave number of the excitation.
At scales larger than the vortex core size ($k\xi\ll1$), such excitations are known as Kelvin waves (KWs) and they play the important role of carrying energy towards the smallest scales of a superfluid \cite{VinenKW}. At such scales, the dynamics of a vortex line can be described by the vortex filament model, according to which the motion of the filament is determined by the self-induced velocity $\mathbf{v}_\mathrm{s.i.}$ of the line on itself \cite{donnelly1991quantized}. . 
This model involves non-local contributions and a singular integral that needs to be regularized \cite{SchwartzVFM}. Note that this model has also been derived at large scales also in the framework of the GP equation \cite{BustNaz}. The simplest approximation that can be done is the well known Local Induction Approximation (LIA), where only the contribution to $\mathbf{v}_\mathrm{s.i.}$ due to the local curvature at each point of the filament is considered. Such approximation is valid when the curvature is much larger than the vortex core size. The LIA model reads \cite{HamaLIA}
\begin{eqnarray}
\dot{\mathbf{s}}(\zeta,t)= \mathbf{v}_\mathrm{s.i.}(\zeta,t),&&      \mathbf{v}_\mathrm{s.i.}(\zeta,t)=\frac{\Gamma}{4\pi}\Lambda\frac{\partial\mathbf{s}}{\partial \zeta}\times\frac{\partial^2\mathbf{s}}{\partial \zeta^2},
\label{Eq:LIAspace}
\end{eqnarray} 
where $\mathbf{s}(\zeta,t)$ is the curve that parametrizes the filament, $\zeta$ is the arc-length. The parameter $\Lambda>0$ is in principle a non-local operator yielding the correct Kelvin wave dispersion relation. At a first approximation and for the sake of simplicity in analytical treatments, it can be considered as a constant. In the case of small displacements of a straight filament oriented along the $z$-axis, the vortex line can be parametrized as $s(z,t)=s_x(z,t)+is_y(z,t)$. At the leading order \eqref{Eq:LIAspace} reduces to
\begin{eqnarray}
\dot{s}(z,t)=v_\mathrm{s.i.}(z,t),&&v_\mathrm{s.i.}(z,t)=i\frac{\Gamma}{4\pi}\Lambda\frac{\partial ^2 }{\partial z^2}s(z,t).
\label{Eq:LIA}
\end{eqnarray}
The LIA equation (\ref{Eq:LIA}) admits solutions in the form of helicoidal waves propagating along the vortex line with a dispersion relation
\begin{equation}
\Omega_\mathrm{LIA}(k)=-\frac{\Gamma\Lambda}{4\pi} k^2.
\label{Eq:OmegaLIA}
\end{equation}
A better description of vortex waves was formally derived from the Euler equations for an ideal incompressible fluid by Sir W. Thomson (Lord Kelvin) \cite{ThomsonKWoriginal} in the case of a hollow vortex, namely if the vorticity is concentrated in a thin tube of radius $a_0$. In this case the frequency of propagation is given by the well known Kelvin wave dispersion relation 
\begin{equation}
\Omega_\mathrm{KW}(k)=\frac{\Gamma}{2\pi a_0^2}\left[ 1 - \sqrt{1 + a_0|k|\frac{K_0(a_0|k|)}{K_1(a_0|k|)}} \right]
\label{Eq:KW}
\end{equation}
where $K_n(x)$ is the modified Bessel function of order $n$ and $a_0$ depends on the model of the vortex core. It has been shown by Roberts \cite{RobertsKW} that the small wave number limit of expression (\ref{Eq:KW}) is valid also for large-scale waves propagating along the superfluid vortex described by the GP equation:
\begin{eqnarray}
\Omega_\mathrm{v}(k)\underset{k\xi\ll 1}{\longrightarrow}\Omega_\mathrm{KW}(ka_0 \rightarrow 0 ) = -\frac{\Gamma}{4\pi}k^2\left(\ln\frac{2}{a_0|k|}-\gamma_\mathrm{E}\right) \nonumber \\,
\label{Eq:KWlim}
\end{eqnarray} 
where $a_0=1.1265\xi$ and $\gamma_\mathrm{E}\sim 0.5772$ is the Euler-Mascheroni constant. On the other hand, at small scales the excitations of a quantum vortex behave as (GP) free particles and the dispersion relation is simply given by \cite{RobertsKW} 
\begin{equation}
\Omega_\mathrm{v}(k)\underset{k\xi\gg 1}{\longrightarrow}-\Omega_{B}({k\xi \rightarrow \infty})=-\frac{\Gamma}{4\pi}k^2,
\label{Eq:freeParticles}
\end{equation}
Note that all the frequencies (\ref{Eq:OmegaLIA}-\ref{Eq:freeParticles}) have an opposite sign with respect to the circulation $\Gamma$, namely KWs rotate opposite to the vortex flow $\mathbf{v}_\mathrm{v}$. 
Since there is not an analytic expression for the full dispersion relation of vortex excitations of the GP model, in the numerics presented in this work we use a fit of the dispersion relation that matches both asymptotic (\ref{Eq:KW}) and (\ref{Eq:freeParticles}). It reads
\begin{eqnarray}
&&\Omega_\mathrm{v}^\mathrm{fit}(k) = \nonumber \\
&&\Omega_\mathrm{KW}(k)\left(1 + \epsilon_\frac{1}{2}(a_0|k|)^\frac{1}{2} + \epsilon_1(a_0|k|) + \frac{1}{2} (a_0|k|)^\frac{3}{2}\right). \nonumber \\
\label{Eq:fit}
\end{eqnarray}
The dimensionless parameters $\epsilon_\frac{1}{2}=-0.20$ and $\epsilon_1=0.64$ are obtained from the measured dispersion relation of a bare vortex tracked in a GP simulation without particles. In Fig.\ref{Fig:fitcomp} the spatiotemporal spectrum of a bare GP vortex  is compared with the result of the fit (solid green line), together with the asymptotics. Note that in Eq. (\ref{Eq:fit}) we used the full Kelvin wave frequency relation (\ref{Eq:KW}) (dashed cyan line) instead of the asymptotic (\ref{Eq:KWlim}) (dotted yellow line). This is because its large $k$ limit $\Omega_\mathrm{KW}(k) \sim \frac{\Gamma}{2\pi a_0^2}(a_0 |k|)^{\frac{1}{2}}$ can be straightforwardly adjusted to obtain the free particle dispersion relation (\ref{Eq:freeParticles}) (dash-dotted magenta line).

\section{Motion of particles trapped by quantum vortex}

We are interested in the behavior of particles captured by quantum vortices. Since hydrogen and deuterium particles used to visualize vortices in superfluid helium experiments are considerably larger than the vortex core (typically $\Rp\sim10^4\xi$) they could be captured not by an isolated vortex but by bundles of many polarized vortices. In such complex system, the large particle size and inertia might affect the vortex dynamics. It is then natural to try to understand how the dynamics of vortices is modified by the presence of the particles, or in other terms, how well particles track superfluid vortices. 

An amazing experimental evidence is that trapped particles distribute themselves at an almost equal spacing (see for instance Ref. \cite{FondaKWExp}). In this work we do not address the physical origins of this distribution, but we adopt it as a hypothesis for setting the initial condition of our simulations. 

We start our discussion by presenting the settings of the GP-P model in our simulations. The GP-P equations are integrated in a 3D periodic domain of dimensions $L_\perp\times L_\perp\times L_\parallel$. The initial conditions consist in a perturbed straight vortex containing small amplitude vortex excitations. The vortex is loaded with a number of particles and then evolved under GP-P dynamics. The computational domain contains three other image vortices in order to preserve periodicity. Only one vortex contains particles whereas the three other are bare. We have used resolutions up to $256\times 256\times 1024$ and $512^3$ collocation points. We express the particle mass as $\Mp=\mathcal{M}M_\mathrm{p}^0$, where $M_\mathrm{p}^0$ is the mass of the displaced superfluid. Therefore, light, neutral and heavy particles have $\mathcal{M}<1$, $\mathcal{M}=1$ and $\mathcal{M}>1$ respectively. Lengths are expressed in units of $\xi$, times in units of $\tau=\xi/c$ and velocities in units of $c$. Further details on the numerical implementation are given in Appendix A.

Figure \ref{Fig:3Dsketch} displays the four different configurations studied in this work. 
\begin{figure}[ht]
\centering
\includegraphics[width=.99\linewidth]{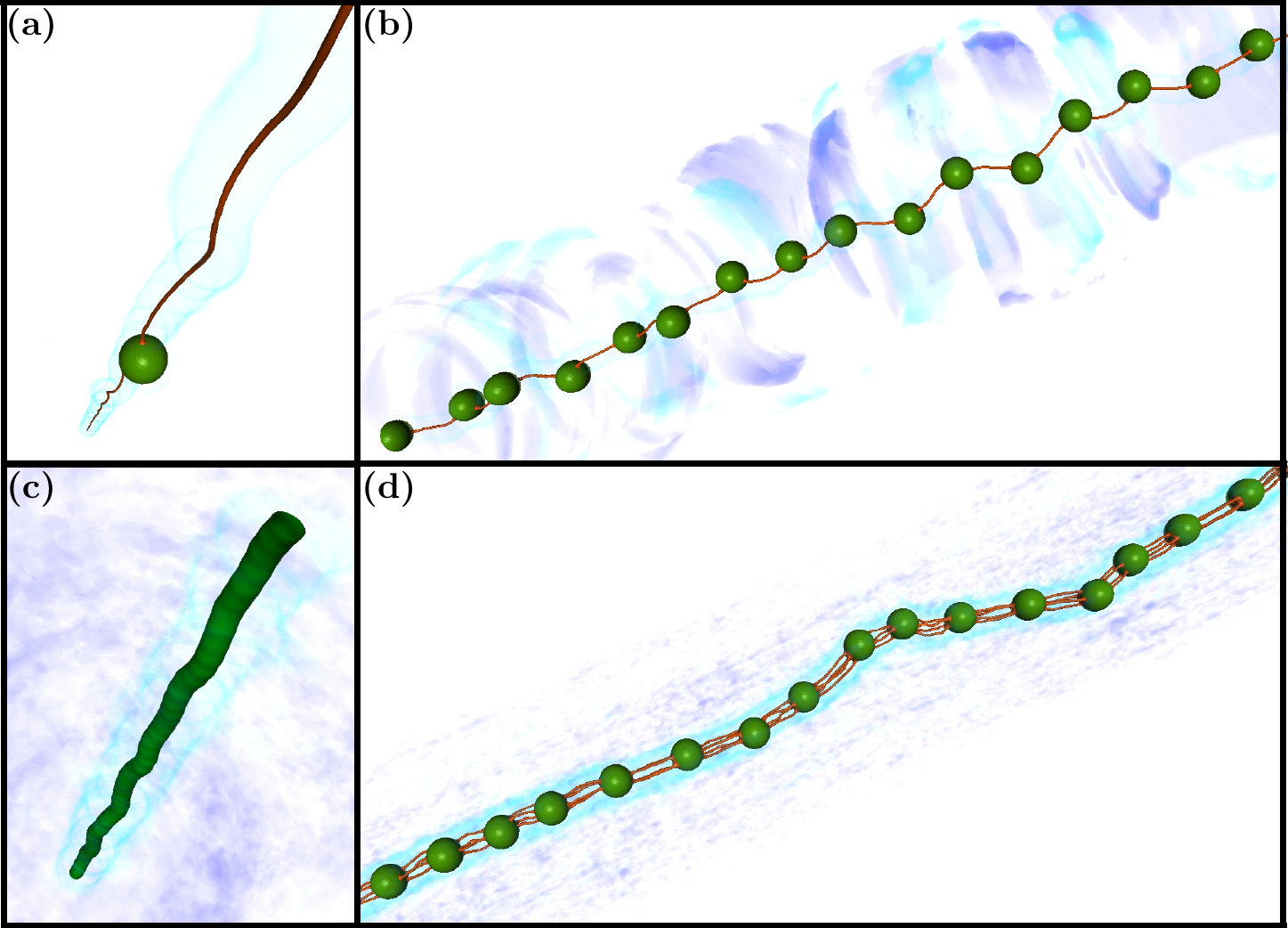}
\caption{Visualisation of particles trapped by superfluid vortices from GP simulations. Vortices are displayed in red, particles in green and sound waves are rendered in blue. \textbf{(a)} A single particle of size $a_\mathrm{p}= 13.1\xi$ trapped in a vortex filament. \textbf{(b)} An array of particles of size $a_\mathrm{p}= 13.1\xi$ and relative distance $d=51.2\xi$. \textbf{(c)} A wire made of $50$ overlapping particles of size $2.7\xi$ trapped in a vortex filament. \textbf{(d)} An array of particles of size $a_\mathrm{p}= 13.1\xi$ trapped in a bundle of 4 vortex filaments. Movies of the simulations can be found in the Supplemental Material.}
\label{Fig:3Dsketch}
\end{figure}
Figure \ref{Fig:3Dsketch}.a shows one particle moving in a quantum vortex which clearly induces KWs on the filament. Figure \ref{Fig:3Dsketch}.b displays an array of particles initially set at equal distances. We have checked that provided that particles are distant enough, they remain equally distributed along the vortex, with very small fluctuations along its axis. Figure \ref{Fig:3Dsketch}.c displays a snapshot in the case where particles strongly overlap creating an almost continuous distribution of mass inside the vortex. Producing this state is possible by properly adjusting the repulsive potential $V_\mathrm{rep}^{ij}$ in Eq. (\ref{Eq:GP}). The purpose of studying this configuration is two-fold. First, from the theoretical point of view it will provide an easier way to describe the role of the particle mass in the vortex dynamics and its effect on vortex excitations. On the other hand, such setting is similar to recent experiments that study the nanowire formation by gold nano-fragments coalescence on quantum vortices \cite{NanoGold2} or experiments with vibrating wires inside quantum vortices in superfluid $^3\mathrm{He}$ and $^4\mathrm{He}$ \cite{ZieveWireHe3,ZieveWireHe4}. Finally, Fig.\ref{Fig:3Dsketch}.d displays a bundle of four equally charged vortices loaded with an array of particles. In all cases, we clearly see the interaction between particles and vortices producing sound (phonon) and Kelvin waves.  Movies of the simulations can be found in the Supplemental Material.

\subsection{Natural frequency of particles trapped by superfluid vortices}

We first consider the dynamics of a particle trapped by an almost straight superfluid vortex. At the leading order this is the classical hydrodynamical problem of a moving sphere with non-zero circulation in an ideal fluid. The main force acting on the particle is the Magnus force, that arises from the pressure distribution generated at the boundary of the particle in such configuration \cite{BatchelorClassic,magnus_sphere}. We introduce the complex variable $q(t)=q_x(t)+iq_y(t)$ for the center of the particle in the plane orthogonal to the vortex filament, and $v=v_x+iv_y$ for the velocity of the ambient superfluid flow. In these variables, the equation of motion for the particle in absence of any external force is \cite{magnus_sphere}
\begin{equation}
\ddot{q}(t)=i\Omega_\mathrm{p}\left(\dot{q}(t)-v\right),\qquad \Omega_\mathrm{p}=\frac{3}{2}\frac{\rho \Gamma a_\mathrm{p}}{M^\mathrm{eff}_\mathrm{p}},
\label{Eq:Magnus}
\end{equation}
where $M_\mathrm{p}^\mathrm{eff}=M_\mathrm{p}+\frac{1}{2}M_\mathrm{p}^0=(\mathcal{M}+\frac{1}{2})M_\mathrm{p}^0$ is the effective mass of the particle and $M_\mathrm{p}^0=\frac{4}{3}\pi \rho a_\mathrm{p}^3$ is the displaced mass of the fluid. In equation (\ref{Eq:Magnus}), the fluid is assumed to be incompressible with density $\rho\sim\rho_\infty$, which is a good approximation when the particle size is larger than the healing length.
From (\ref{Eq:Magnus}) we can derive the temporal spectrum of the particle position 
\begin{equation}
|\hat{q}(\omega)|^2=\frac{\Omega_\mathrm{p}^2|\hat{v}(\omega)|^2}{\omega^2(\omega-\Omega_\mathrm{p})^2}
\label{Eq:qtilde}
\end{equation}
where $\hat{q}(\omega)=\int q(t)e^{-i\omega t}\,\mathrm{d}t$ and $\hat{v}(\omega)=\int v(t)e^{-i\omega t}\,\mathrm{d}t$. The vortex line tension, which is responsible for the propagation of Kelvin waves \cite{sonin_book}, is implicitly contained in the superfluid flow $v$ in Eq. (14). It generates particle oscillations in the rotation direction opposite to the flow generated by the vortex. However, from Eq. (15) we see that the particle motion is dominated by a precession with frequency $\Omega_\mathrm{p}$, which has the same sign of $\Gamma$ and therefore has the same direction of the vortex flow. Such frequency is the natural frequency of the particle: expressing it as a function of $\mathcal{M}$ we get:
\begin{equation}
\Omega_\mathrm{p}=\frac{9}{4\pi}\frac{\Gamma}{a_\mathrm{p}^2(2\mathcal{M}+1)}.
\label{Eq:Omega_p_2}
\end{equation}
For current experiments using particles as probes, such characteristic frequency is of order 10--100Hz, which is actually measurable \cite{MatthieuGibert}.

\begin{figure}[ht]
\centering
\includegraphics[width=1.\linewidth]{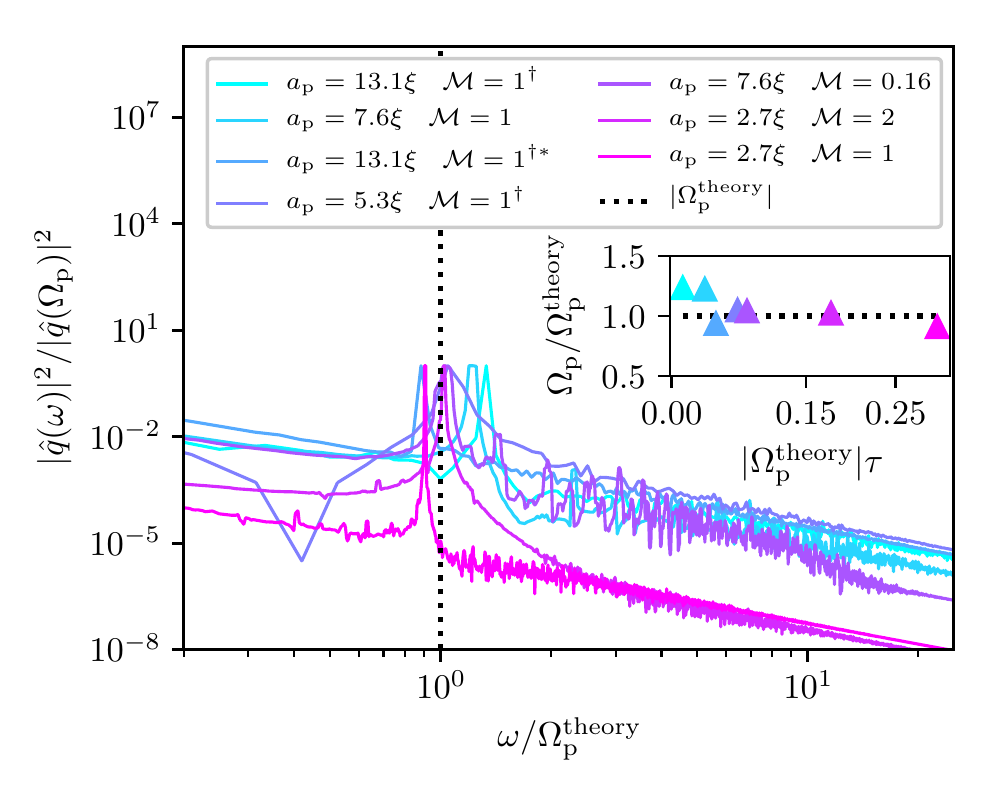}
\caption{Temporal spectra of the particle positions for different values of the natural frequency $\Omega_\mathrm{p}$, obtained varying mass and size of the particles. The expected natural frequency $|\Omega^\mathrm{theory}_\mathrm{p}|$ (\ref{Eq:Omega_p_2}) is the dotted vertical line. \textbf{Inset}:  Comparison of the measured natural particle frequency with the theory.  $^\dagger$: the particle considered belongs to a particle array. $^*$: the particle considered is trapped in a bundle of 4 vortices.}
\label{Fig:ParticleFrequency}
\end{figure}

We have performed a series of numerical experiments with particles trapped in a superfluid vortex excited with small amplitude Kelvin waves. Measurements of temporal spectra (\ref{Eq:qtilde}) for particles characterized by different values of $\Omega_\mathrm{p}$ are reported in Fig.\ref{Fig:ParticleFrequency}. In the x--axis of the plot we have the angular frequencies with the same sign of $\Gamma$.
The different natural frequencies have been obtained varying the mass and the size of the particles. The observed peak at $\Omega_\mathrm{p}$ is well predicted by Eq. (\ref{Eq:qtilde}). The natural frequency is also observed for particles in the particle-array configuration. In particular, if particles are attached to a bundle of $N_\mathrm{v}$ quantum vortices instead of a single filament, the corresponding characteristic frequency is $N_\mathrm{v}$ times larger. The case of a bundle of $N_\mathrm{v}=4$ is also reported in Fig.\ref{Fig:ParticleFrequency}, in a remarkable agreement with theory. This has an important experimental implication. Measuring the natural frequency $\Omega_\mathrm{p}$  could give an independent estimate of the circulation (and therefore of the number of vortices) in the bundles visualized by the particles in superfluid helium experiments.

Note that in general the vortex line tension could have a non-trivial coupling with the particles and lead to a modification of the precession frequency $\Omega_\mathrm{p}$. Indeed, in the idealized derivation of Eq. (\ref{Eq:Magnus}), it is assumed that the particle center coincides with the center of a straight vortex line.  In principle, one should solve Eq. (\ref{Eq:Magnus}) together with the equation of motion of the vortex, taking into account the proper boundary conditions between a sphere and a vortex filament \cite{SchwartzVFM}, that will include restoring forces maintaining the particle trapped.  Accounting for such phenomena might lead to a more accurate prediction of the precession frequency. However, the GP system naturally contains all these effects. Therefore, given the agreement between the prediction (\ref{Eq:Omega_p_2}) and GP numerical simulations, we conclude that the modification of the particle natural frequency $\Omega_\mathrm{p}$ due to the coupling at the particle-vortex boundary is a negligible effect. The simple formula (\ref{Eq:Omega_p_2}) can be thus safely used as a first estimate in current experiments.
 
\subsection{Dispersion relation of a massive quantum vortex}
As already mentioned above, in order to study the dynamics of an array of particles and their interaction with vortex waves in a setting like Fig.\ref{Fig:3Dsketch}.b or Fig.\ref{Fig:3Dsketch}.d,  It is instructive to first analyze the case of a massive quantum vortex, as the one in Fig.\ref{Fig:3Dsketch}.c. Our considerations are necessary to give a picture of the role of inertia in the propagation of vortex wave excitations. They are not meant to model a real wire, for which some results are well known in literature \cite{VinenMass,FetterElasticWaves} and has been used to measure the quantized circulation in superfluid helium \cite{KarnQ,DavisQ}. We consider a wire of length $L_\mathrm{w}$, radius $a_\mathrm{w}$ and mass $M_\mathrm{w}$, filling a superfluid vortex.
The effective mass is $M^\mathrm{eff}_\mathrm{w}=M_\mathrm{w} + M_\mathrm{w}^0$ and the displaced mass is now $M_\mathrm{w}^0=\rho L_\mathrm{w}\pi a_\mathrm{w}^2$. Since such wire possesses a circulation, each mass element is driven by Magnus force as in Eq. (\ref{Eq:Magnus}), but with a different prefactor  \cite{BatchelorClassic}
\begin{equation}
\Omega_\mathrm{w} = \frac{\rho\Gamma L_\mathrm{w}}{M^\mathrm{eff}_\mathrm{w}},
\label{Eq:prefactor_cilindro}
\end{equation}
which arises because of the geometrical difference between a spherical particle and a cylinder. We allow the wire to deform, that means that the complex variable $q$ is now a function of the $z$ component too. Such physical system is analogous to a massive quantum vortex with a finite size core, which is already well known in literature \cite{VinenMass,FetterElasticWaves}, and it has been used to measure the quantized circulation in superfluid helium \cite{KarnQ,DavisQ}. If the curvature radius is much greater than the wire radius and the healing length, the flow velocity $v$ can be approximated by the self-induced velocity of the vortex filament on itself. In the LIA approximation, the self-induced velocity is simply given by $v_{\rm s,i}$ in Eq. (\ref{Eq:LIA}). The dynamics of the wire is therefore driven by the equation
 \begin{equation}
 \ddot{q}(z,t)=i\Omega_\mathrm{w}\left(\dot{q}(z,t)-i\frac{\Gamma}{4\pi}\Lambda\frac{\partial ^2 }{\partial z^2}q(z,t)\right).
 \label{Eq:Magnus_rope}
 \end{equation}
In this simplified model, we are neglecting modes propagating along the wire due to elastic tension and the wave number dependence of the added mass. This choice is done because we want to focus on the inertial effects that will be relevant in the case of a particle array, developed in the following section. Equation (\ref{Eq:Magnus_rope}) allows as solution linear circularly polarized waves in the form $q(z,t)=q_0e^{i(\Omega_\mathrm{M}^{\pm} t - kz)}$, where the frequency is given by 
\begin{equation}
\Omega_\mathrm{M}^{\pm}(k) = \frac{\Omega_\mathrm{w}}{2} \pm \frac{1}{2}\sqrt{\Omega_\mathrm{w}^2 + \frac{\Omega_\mathrm{w}\Gamma\Lambda}{\pi}k^2}.
\label{Eq:MagnusLIA}
\end{equation}
More generally, one can consider a phenomenological extrapolation based on a more realistic model for the self-induced velocity of the vortex in the equation (\ref{Eq:Magnus_rope}), so that the dispersion relation of waves propagating along the wire is generalized as
\begin{equation}
\Omega_\mathrm{M}^{\pm}(k) = \frac{1}{2}\left[\Omega_\mathrm{w} \pm \sqrt{\Omega_\mathrm{w}^2-4\Omega_\mathrm{w}\Omega_\mathrm{v}(k)} \right],
\label{Eq:omega_Magnus}
\end{equation}
where $\Omega_\mathrm{v}(k)$ is the \emph{bare} vortex wave frequency and depend on the model chosen for the self-induced velocity. We will refer to (\ref{Eq:omega_Magnus}) as the ``massive vortex wave'' dispersion relation. In the LIA approximation we have $\Omega_\mathrm{v}(k)=\Omega_\mathrm{LIA}(k)$ (\ref{Eq:OmegaLIA}) and we recover Eq. (\ref{Eq:MagnusLIA}), but a more accurate result is expected if the wave propagation is instead described by $\Omega_\mathrm{KW}(k)$ or by the measured dispersion relation $\Omega_\mathrm{v}^{\rm fit}(k)$ (\ref{Eq:fit}). Note that the zero-mode of the branch $\Omega_\mathrm{M}^+$ coincides with $\Omega_\mathrm{w}$ and does not vanish even if $M_\mathrm{w}=0$ because of the added mass $M_\mathrm{p}^0$. This is related to the fact that the wire possesses an effective inertia because during its motion it has to displace some fluid \cite{VinenMass,SimulaMass}. In the limit $k\xi\ll 1$, the result (\ref{Eq:omega_Magnus}) can be obtained from the one derived in Ref. \cite{FetterElasticWaves} using fluid dynamic equations to study ions in superfluid helium. 

\begin{figure}[ht]
\centering
\includegraphics[width=.99\linewidth]{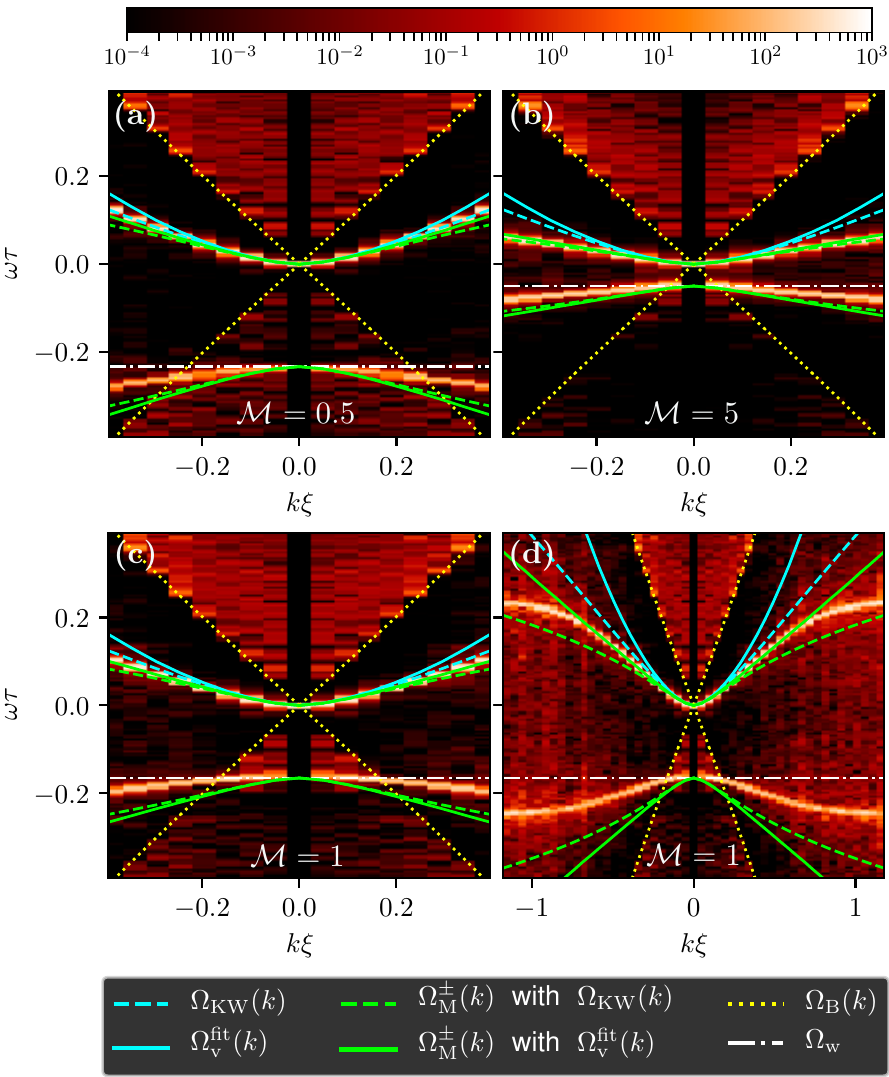}
\caption{Spatiotemporal spectra of massive vortices for different masses. The vortex length is $L_\mathrm{w}=128\xi$ and there are $N_\mathrm{p}=50$ particles of radius $a_\mathrm{p}=2.7\xi$, with repulsion radius $r_0=2L_\mathrm{w}/(N_\mathrm{p}a_\mathrm{p})$. Dotted yellow line is the Bogoliubov dispersion relation $\Omega_\mathrm{B}(k)$ (\ref{Eq:Bogoliubov}). Dashed cyan line is low-$k$ KW dispersion relation $\Omega_\mathrm{KW}(k)$ (\ref{Eq:KW}). Solid cyan line is full fitted vortex wave dispersion relation $\Omega_\mathrm{v}^\mathrm{fit}(k)$ (\ref{Eq:fit}). Dash-dotted green lines are massive vortex wave dispersion relation $\Omega_\mathrm{M}(k)$(\ref{Eq:omega_Magnus}) computed using low-$k$ KW dispersion relation. Solid green lines are massive vortex wave dispersion relation computed using full fitted vortex wave dispersion relation. Dotted horizontal white line is the natural frequency $\Omega_\mathrm{w}$ (\ref{Eq:prefactor_cilindro}). The other parameters of the simulations are $L_\perp=L_\parallel = 128\xi$ and $N_\perp = N_\parallel = 256$. \textbf{a)} $\mathcal{M}=0.5$; \textbf{b)} $\mathcal{M}=5$; \textbf{c)} $\mathcal{M}=1$; \textbf{d)} Same as \textbf{c)}, but displaying the full range.}
\label{Fig:SpectraCyl}
\end{figure}

We build numerically a massive vortex placing a large number of small overlapping particles along a vortex filament. We set the repulsion between particles at a radius $r_0=2L_\mathrm{w}/(N_\mathrm{p}a_\mathrm{p})$ (see Appendix A), so that they are kept at constant distance $r_0/2$. Such system mimics a continuum of matter with total mass given by the sum of all particle masses $M_\mathrm{w}=N_\mathrm{p}M_\mathrm{p}=N_\mathrm{p}M_\mathrm{p}^0\mathcal{M}$. We have checked that the repulsion among particles leads to matter sound waves with frequencies that are sub-leading with respect to other terms present in \eqref{Eq:Magnus_rope}. We initially excite the system with small amplitude Kelvin waves and we let it evolve under GP-P dynamics. Figure \ref{Fig:3Dsketch}.c shows a typical snapshot of the system but in the case of a larger initial perturbation (in order to enhance visibility). We then use the particle positions to construct the spatiotemporal spectrum $S_q(k,\omega)\sim |\hat{q}(k,\omega)|^2$, with $\hat{q}(k,\omega)$ the time and space Fourier transform of $q(z,t)$ (see Appendix B for further details). Density plots of $S_q(k,\omega)$ are displayed in Fig.\ref{Fig:SpectraCyl} for different values of the particle mass. For a better presentation, we have chosen $\Gamma<0$ so that vortex wave frequencies lay in the upper plane. This convention will be adopted also in the following section.

We first observe that the massive vortex is able to capture the Bogoliubov dispersion relation $\Omega_\mathrm{B}(\mathbf{k})$ (\ref{Eq:Bogoliubov}) due to the presence of excitations in the superfluid, as displayed by yellow dotted lines in Fig.\ref{Fig:SpectraCyl}. The bare Kelvin wave dispersion relation $\Omega_\mathrm{KW}(k)$ and the measured bare vortex frequency spectrum $\Omega_\mathrm{v}^{\rm fit}(k)$ are displayed by the cyan dashed and solid lines respectively. They coincide in the limit $k\xi\ll 1$, as expected. The corresponding massive vortex wave predictions (\ref{Eq:omega_Magnus}) are also displayed in green dashed and solid lines. For low masses, the effect of inertia is negligible, so that massive vortex wave (\ref{Eq:omega_Magnus}) and bare vortex wave (\ref{Eq:fit}) predictions are similar. As the mass increases, the wire inertia becomes important and the measured frequencies of the wire excitations decrease at small scales, in good agreement with the massive vortex wave prediction. The model (\ref{Eq:omega_Magnus}) is not expected to give a good explanation for the negative branches, as it neglects the details of the internal structure of the wire, as well as the dependence on the wave number of the effective mass. Such features, that are out of the scope of the present work, are taken into account in Ref. \cite{FetterElasticWaves} in the case of an elastic and massive hollow vortex (with no notion of the free-particle behavior of vortex excitations at small scales). The predicted natural frequency of the wire $\Omega_{\rm W}=| \Omega^+_{\rm M}(0)|$ is clearly reproduced by the numerical measurements and it does not become infinite when $\mathcal{M}\to0$ because of the added mass effect. For completeness, Fig.\ref{Fig:SpectraCyl}.d displays the dispersion relation over the full accessible range of wave numbers. The dispersion curves are bent due to the discreetness of the wire at scales of order $k\xi\sim 0.8$. 
Note that the KW dispersion relation (dashed cyan line) seems to be very similar to the fitted one (solid cyan line). However, the difference between the two is apparent in Fig.\ref{Fig:SpectraCyl}.d. Moreover, it is clear how the massive vortex wave dispersion relation computed using  $\Omega_\mathrm{v}(k)=\Omega_\mathrm{v}^\mathrm{fit}(k)$ (solid green line) fits the data for all the masses analyzed. In particular, in Fig.\ref{Fig:SpectraCyl}.d, it is shown that it can predict the dispersion relation of a massive vortex wire with relative mass $\mathcal{M}=1$ up to a wave number $k\xi\sim 0.7$. This is not the case for the massive vortex wave dispersion relation computed using  $\Omega_\mathrm{v}(k)=\Omega_\mathrm{KW}(k)$ (dashed green line).
We thus conclude that the main effect of the inertia of the particles constituting the wire is to modify the frequency spectrum of vortex wave, as follows from simple hydrodynamical considerations.

\subsection{Frequency gaps and Brillouin zones for an array of trapped particles}
\label{sec:brillouin}
\begin{figure}[h!]
\includegraphics[width=.99\linewidth]{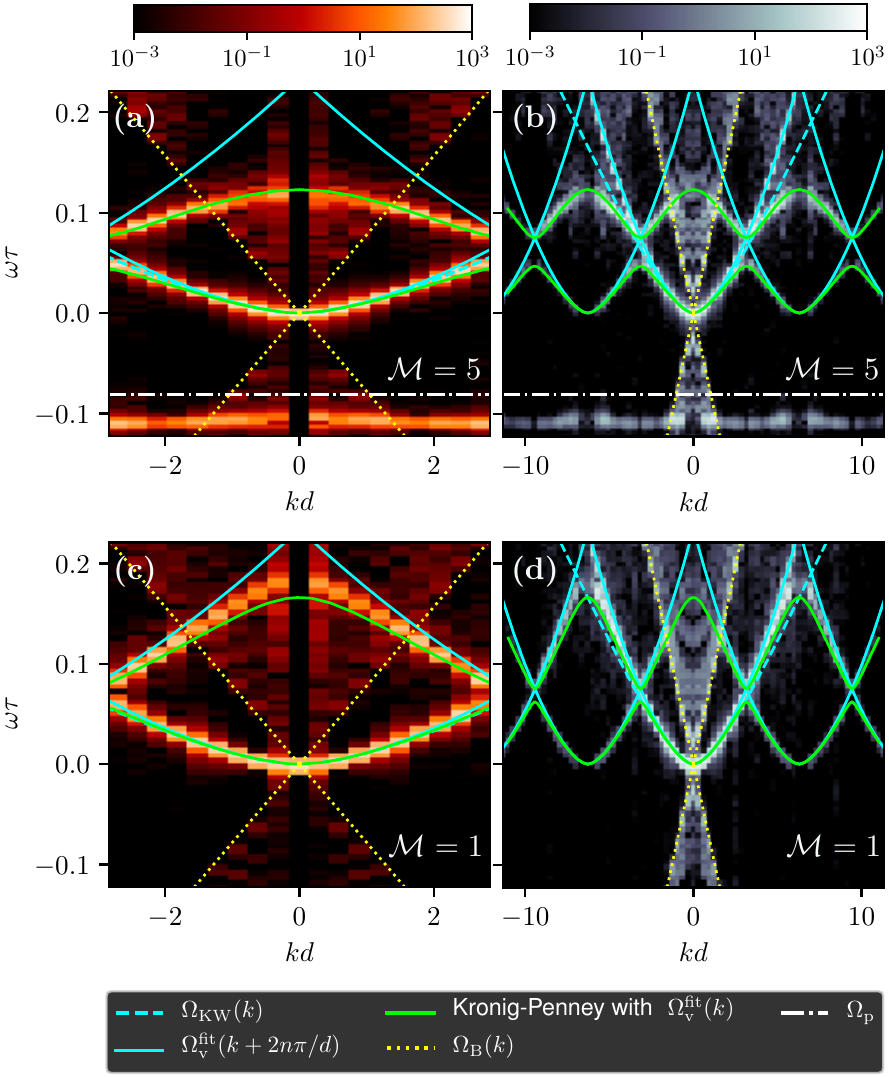}
\caption{Spatiotemporal spectra computed from the particle positions (left) and from the wave function $\psi$ (right) for an array of particles with mass $\mathcal{M}=5$ (up) and $\mathcal{M}=1$ (down) . Solid green lines are the contour-plot of the dispersion relation (\ref{Eq:KP}) computed with $\Omega_\mathrm{v}^\mathrm{fit}$ (\ref{Eq:fit}). Dashed cyan line is low-$k$ KW dispersion relation $\Omega_\mathrm{KW}(k)$ (\ref{Eq:KW}). Solid cyan line is the fitted vortex wave dispersion relation (\ref{Eq:fit}). Dotted yellow line is Bogoliubov dispersion relation $\Omega_\mathrm{B}(k)$ (\ref{Eq:Bogoliubov}). Dash-dotted horizontal white line is the predicted natural frequency $\Omega_\mathrm{p}$. 
The other parameters of the particles are $d=12.8\xi$, $a_\mathrm{p}=2.7\xi$, $r_0 = 4a_\mathrm{p}$.
The size of the computational box is $L_\perp=L_\parallel = 256\xi$, with $N_\perp = N_\parallel = 512$ collocation points.
} 
\label{Fig:KP}
\end{figure}

Now we shall address the main question of this work. How well do particles, seating in a quantum vortex, track vortex waves? In order to study this problem, we consider an array of particles as the one displayed in Fig.\ref{Fig:3Dsketch}.b. Particles are placed in a quantum vortex, initially separated by a distance $d$. The system is excited by superimposing small amplitude KWs. We can build a discrete spatio-temporal spectrum $S_q(k,\omega)$ of the measured vortex excitations by using the displacement of particles in the plane perpendicular to the vortex. In Fig.\ref{Fig:KP}.a and Fig.\ref{Fig:KP}.c we display the particle spatio-temporal spectra for an array of $N_\mathrm{p}=20$ particles of size $\Rp=2.7\xi$ with masses $\mathcal{M}=5$ and $\mathcal{M}=1$ respectively, placed at a distance $d=12.8\xi$.
The Bogoliubov waves are still weakly sampled by the particles, as displayed by yellow dotted lines. Surprisingly, a higher frequency branch appears. Such pattern is similar to those observed in the typical energy spectra of crystals \cite{kittel1976introduction}. Particles are actually able to sample the vortex excitations only in the first Brillouin zone, namely they cannot see wave numbers larger than $\pi/d$. However, spatio-temporal spectra can be also computed by directly using the superfluid wave function. Performing the time and space Fourier transform of $\psi$ we define the spectrum $S_\psi(k,\omega)=|\hat{\psi}(k_x=0,k_y=0,k,\omega)|^2$. The corresponding spectra $S_\psi$ are shown in Fig.\ref{Fig:KP}.b and Fig.\ref{Fig:KP}.d where wave numbers go now up to $kd\sim10$, giving access to all the small scales solved by the numerical simulations. Several Brillouin zones are clearly appreciated, as well as the opening of band gaps in the dispersion relation. At the same time, Bogoliubov modes can be observed and also bare vortex waves. The latter belong to the image vortices in the computational domain, where no particles have been attached.

\begin{figure}[h!]
\centering
\includegraphics[width=.99\linewidth]{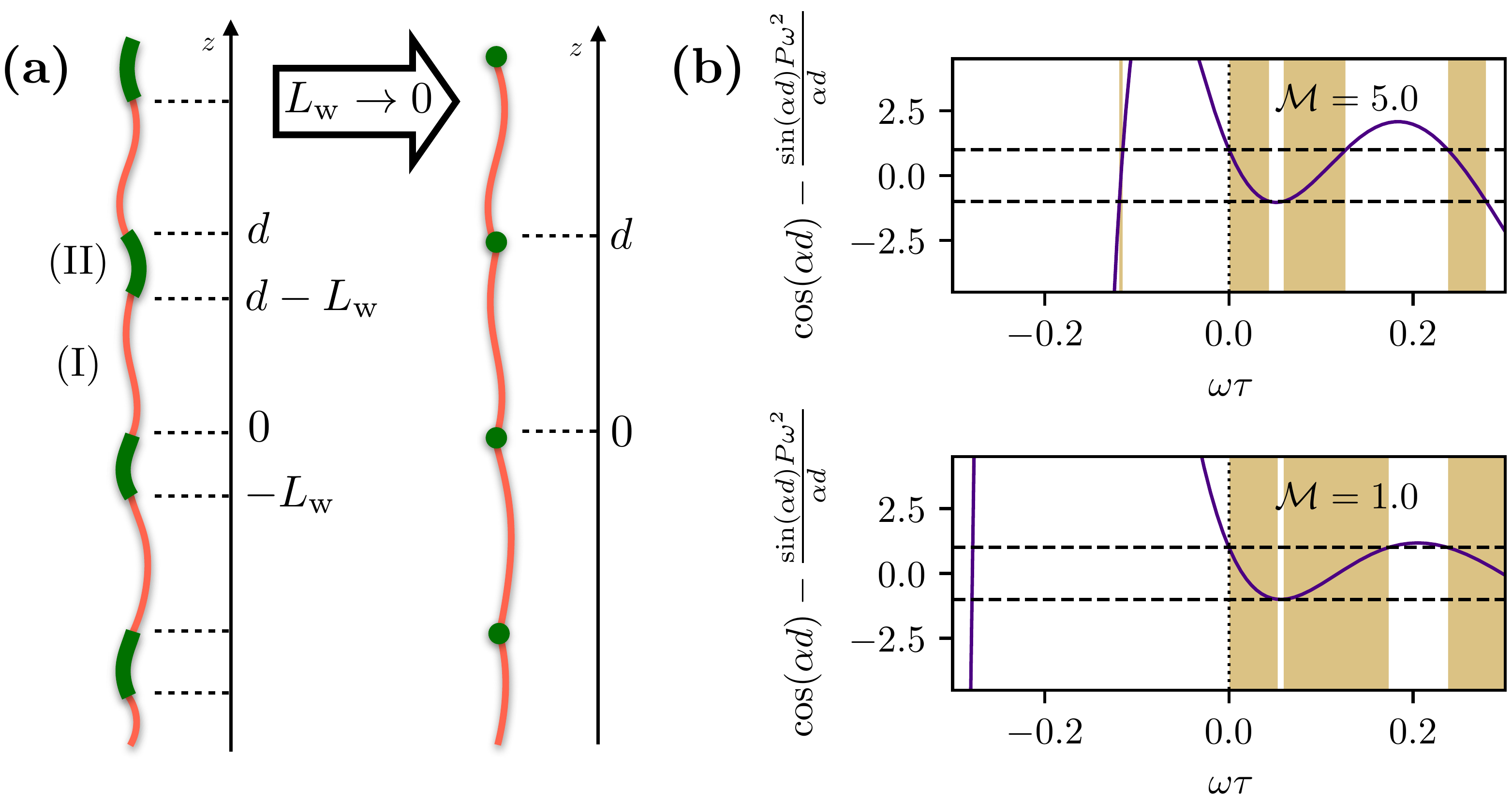}
\caption{\textbf{a)} Sketch of the lattice vortex wave model. Bare vortex segments are in red and massive vortex segments are in green. \textbf{b)} r.h.s. of Eq. (\ref{Eq:KP}) computed with LIA as a function of $\omega\tau$ for an array of particles with radius $a_\mathrm{p}=2.7\xi$ and mass $\mathcal{M}=5$. .  Bands of allowed frequencies are displayed in gold.  \textbf{c)} The same of \textbf{b)} but for particles with mass $\mathcal{M}=1.$}
\label{Fig:Sketch}
\end{figure}

The presence of particles clearly affects the propagation of waves along the vortex line inducing high frequency excitations not only for small but also for large wave lengths. The intuitive idea is that when a vortex wave reaches a particle, it is partially reflected or transmitted, depending on the mass and the size of the particles, and eventually on its own frequency. This reminds us of the standard quantum-mechanical problem of an electron described by the (linear) Schr$\mathrm{\ddot{o}}$dinger equation hitting a potential barrier. Furthermore, if particles are set at almost equal distances, the system is similar to an electron propagating in a periodic array of potential barriers, as in the Kronig-Penney model \cite{KronigPenney,kittel1976introduction}. In order to apply quantitatively this intuition and explain the opening of band gaps in the dispersion relation of vortex wave excitations, we start by considering an artificial system made of segments of bare quantum vortex of length $(d-L_\mathrm{w})$, alternated with massive vortex wires of length $L_\mathrm{w}$. A sketch of the problem is given in Fig.\ref{Fig:Sketch}.a.
To recover the excitations in the case of the particle array, we will later take the limit $L_{\rm w}\to0$, keeping the mass of the wires equal to the effective mass of the particles. The resulting effective theory must be intended as an asymptotic limit of the actual system for long waves $ka_\mathrm{p}\ll 1$, in which the nonlinear interactions of the vortex excitations are neglected and the complexity of the vortex--particle boundary is ignored. The accuracy of such model has to be checked by comparing its predictions with the results of the GP simulations. The motion of the bare vortices is driven by the self-induced velocity that leads to the propagation of vortex waves, while the wires are driven by the Magnus force. For the sake of simplicity, we first consider the LIA approximations (\ref{Eq:LIA}) and (\ref{Eq:Magnus_rope}) respectively. The dynamics is thus given in each zone by
\begin{eqnarray}
    \dot{q}(z,t) = i\frac{\Gamma}{4\pi}\Lambda\frac{\partial ^2 }{\partial z^2}q(z,t)	\quad (\mathrm{I})\nonumber \\
    \ddot{q}(z,t) = i\Omega_\mathrm{w}\left[ \dot{q}(z,t) -  i\frac{\Gamma}{4\pi}\Lambda\frac{\partial ^2 }{\partial z^2}q(z,t)\right]	\quad (\mathrm{II}) \nonumber \\
    \label{Eq:system}
\end{eqnarray}
where $(\mathrm{I})$ is the region $0<z<d-L_\mathrm{w}$ and $(\mathrm{II})$ is the region $d-L_\mathrm{w}<z<d$. 
Note that the use of LIA in the system (\ref{Eq:system}) is rather qualitative, given the high level of complexity of the problem. In particular it ignores the nonlocal dynamics of the vortex, does not reproduce the good dispersion relation of vortex excitations and may not be able to take into account the exact boundary condition between the particles and the vortex. However, it allows us to introduce some general physical concepts and perform a fully analitically treatment of the problem. The effective model will be then generalized in order to take into account a more realistic description of vortex waves and provide quantitative predictions.
The dispersion relation can be found borrowing standard techniques from solid state physics, in particular by adapting the solution of the Kronig-Penny model \cite{KronigPenney,kittel1976introduction}. We look for a wave solution $q(z,t)=\Phi(z)e^{i\omega t}$, where the spatial function $\Phi(z)$ can be written in the form $\Phi(z)=e^{ikz}u(z)$ according to Bloch theorem, where $u(z)$ is a periodic function of period $d$ \cite{Floquet1883}. The key point is the imposition of continuity and smoothness of the function $\Phi(z)$ as well as periodicity of $u(z)$ and its derivative. These constrains lead to an implicit equation relating the frequency of the excitations $\omega$, the wavenumber $k$ and all the physical parameters. The full derivation is explained in Appendix C. The last step in order to describe the excitations of the particle array, is to take the limit $L_{\rm w}\to0$ at constant $M_{\rm eff}$. The dispersion relation is finally determined by the implicit equation
\begin{equation}
\cos(kd) = \cos(\alpha_\omega d) - \frac{\sin(\alpha_\omega d)}{\alpha_\omega d}P\omega^2, \label{Eq:KP}
\end{equation}
where $P=3\pi d a_\mathrm{p}/\Lambda\Gamma\Omega_\mathrm{p}$ and $\alpha_\omega$ satisfies the equation $\Omega_\mathrm{LIA}(\alpha_\omega)=\omega$:
\begin{equation}
\alpha_\omega= \sqrt{-\frac{4\pi\omega}{\Gamma\Lambda}}.\label{Eq:alphaVsomega}
\end{equation}
In Figures \ref{Fig:Sketch}.b-c the r.h.s. of Eq. (\ref{Eq:KP}) is plotted as a function of $\omega\tau$ for heavy and light small particles (that is low and high $\Omega_{\rm p}$). The curve must be equal to $\cos(kd)$ and this selects the only allowed frequencies (displayed in gold). It is exactly the same mechanism that leads to the formation of energy bands in crystals \cite{kittel1976introduction}. 

The previous calculations can be directly generalized for more realistic wave propagators (see Appendix C). In particular, if we consider a dispersion relation $\Omega_\mathrm{v}(k)$ for the vortex excitations, the only change in the result (\ref{Eq:KP}) is the functional dependence of $\alpha_\omega$ (\ref{Eq:alphaVsomega}), that must satisfy $\Omega_\mathrm{v}(\alpha_\omega)=\omega$. Furthermore, the constant $P$ becomes independent of any adjustable parameter: $P=3\pi d a_\mathrm{p}/\Gamma\Omega_\mathrm{p}$. 
We consider the dispersion relation ${\Omega_\mathrm{v}^{\rm fit}}(\omega)$ (\ref{Eq:fit}) that matches large and and short scales excitations and we invert it numerically to find $\alpha_\omega$.

In Fig.\ref{Fig:KP} the contour-plot of the theoretical prediction \eqref{Eq:KP} obtained this way is compared with the numerical data (solid green lines), exhibiting a remarkable agreement with the observed excited frequencies.  From Fig.\ref{Fig:Sketch}.b, we remark that the only allowed negative frequencies lay a in a thin band around $\Omega_\mathrm{p}$. This is also in qualitative agreement with the data. Note that the bare Kelvin wave dispersion relation (\ref{Eq:KW}) (dashed cyan line), and the fitted bare vortex wave dispersion relation (\ref{Eq:fit}) (solid cyan line) are very similar in Fig.\ref{Fig:KP}. The reason is that the smallest scale that can be solved by the considered array of particles is $k\xi=0.25$ (i.e. $kd=\pi$) and for wave numbers smaller than this value $\Omega_\mathrm{v}^\mathrm{fit}(k)$ tends to $\Omega_\mathrm{KW}(k)$ by construction.

\begin{figure}[h!]
\includegraphics[width=.99\linewidth]{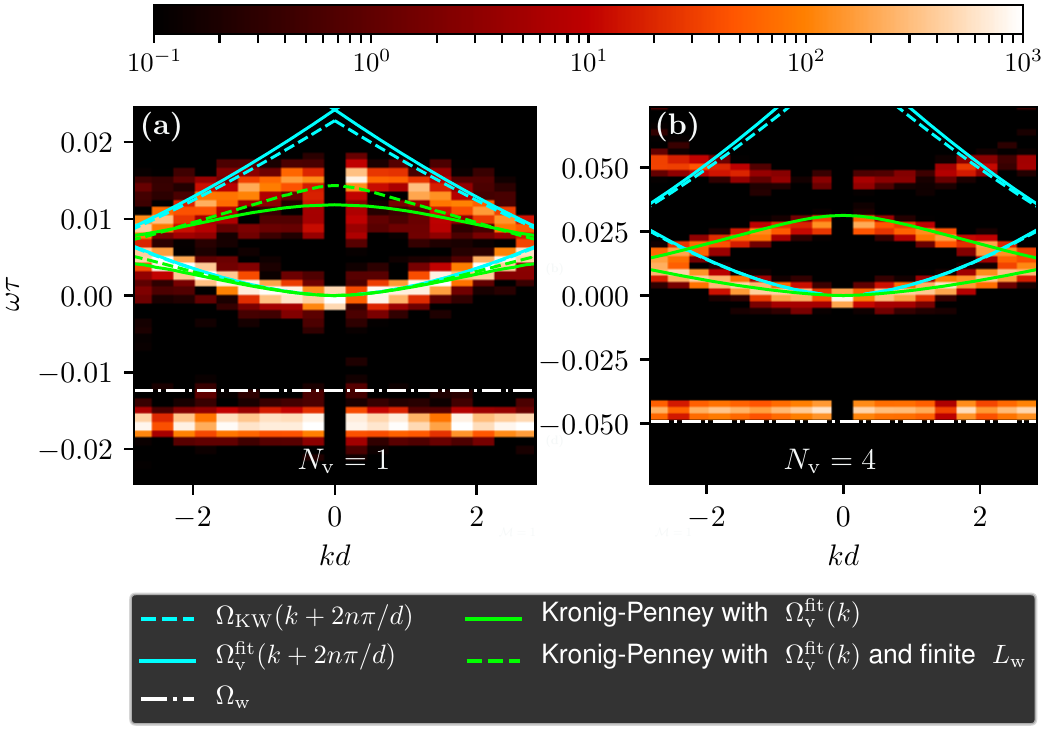}
\caption{Spatiotemporal spectra computed from the particle positions for an array of large particles of radius $a_\mathrm{p}=13.1\xi$ and mass $\mathcal{M}=1$ placed at a distance $d=51.2\xi$. The repulsion radius is $r_0=2a_\mathrm{p}$. Solid green lines are the contour-plot of the dispersion relation (\ref{Eq:KP}) computed with $\Omega_\mathrm{v}^\mathrm{fit}$. Dashed cyan line is low-$k$ KW dispersion relation $\Omega_\mathrm{KW}(k)$. Solid cyan line is the fitted vortex wave dispersion relation. Dash-dotted horizontal white line is the predicted natural frequency $\Omega_\mathrm{p}$. 
\textbf{(a)} Particles set in a single vortex. Dashed greed line is the dispersion relation (\ref{Eq:FullKP}) before the limit computed using a finite $L_{\rm w}=2\Rp$. For the LIA calculations $\Lambda=2.6$.
\textbf{(b)} Particles set in a bundle of $4$ vortices. The dispersion relation (\ref{Eq:KP}) has been computed using $\Omega_\mathrm{v}^\mathrm{fit}$ with an effective core size of $a_0=12\xi$ (see text). 
The other parameters of the simulations are $L_\perp=1024\xi$, $L_\parallel = 256\xi$ and $N_\perp = 1024$, $N_\parallel = 256$.} 
\label{Fig:KPbig}
\end{figure}

In order to make a closer connection with experiments, we now describe an array of larger particles of size $\Rp=13.1\xi$ and relative mass $\mathcal{M}=1$ set in a single quantum vortex and in a bundle of composed of four vortices. The corresponding spatio-temporal spectra $S_{\rm p}(k,\omega)$ are displayed in Fig.\ref{Fig:KPbig}.
In principle such setting should not be well described by our theoretical approach. However, the excitation curves can be reproduced by using the model before the limit $L_\mathrm{w}\rightarrow 0$ (\ref{Eq:FullKP}) and phenomenologically replacing $L_{\rm w}= 2\Rp$ while keeping $\Omega_\mathrm{w}^\mathrm{eff} = \Omega_\mathrm{p}^\mathrm{eff}$. The agreement is remarkably good, considering the rough modeling that has been done. The case of a bundle in Fig.\ref{Fig:KPbig}.b is even more striking. At large scales, we could expect that such system is analogous to a hollow vortex with four quanta of circulation and some effective core size. We have estimated the effective core size by measuring the mean distance between the vortices. The theoretical prediction (\ref{Eq:KP}) combined with this phenomenological approach is still impressively matching the numerical data.

\section{Discussion}\label{sec:conclusion}

In this work we have presented a theoretical and numerical study of the interaction between quantum vortices and a number of particles trapped in it. We have first pointed out that a trapped particle oscillates with a well defined natural frequency that depends on its mass and the circulation of the flow surrounding it. Because of the typical values of particle parameters used in current superfluid helium experiments, such frequency should be measurable. This measurement can thus provide an independent way of estimating the number of vortices constituting the bundles at which particles are attached.  

Based on the experimental evidence that particles spread along quantum vortices keeping a relatively constant inter-particle distance, we have studied how the particles modifies the vortex excitations. The most exciting result of this work is the strong analogy with solid state physics. Here, particles play the role of ions in the periodic structure of a crystal and vortex excitations that of the electrons. When an electron propagates, it feels the ions as the presence of a periodic array of potential barriers. One of the simplest and idealized descriptions of this physical phenomenon is the Kronig-Penney model, where the barriers have a constant height $U_0$. Similarly, vortex waves propagate and interact with particles and we have shown that a similar theoretical approach can be used. The main difference is that the constant height of the barriers in the standard Kronig-Penney model induces constant shift of the energy (frequency here). As a consequence, the lowest energy level in a crystal is different from zero (unlike the case of free electrons). Instead, in the vortex case, the interaction potential is due to Magnus force and depends on the frequency. Comparing the models, we can then establish a mathematical analogy (see Eq. (\ref{Eq:KP}) and Ref.\cite{KronigPenney,kittel1976introduction}) by noticing that the effective potential in the case of vortex excitations is given by
\begin{equation}
U_0\sim \omega^2/\Omega_{\rm p} \propto \omega^2 M_\mathrm{p}^\mathrm{eff}
\end{equation}
The height of the potential is thus proportional to the squared frequency of the incoming wave and to the particle mass. In particular, for very low frequencies the presence of particles does not perturb much the vortices and large scale Kelvin waves could be tracked by directly measuring the particle dynamics. Moreover, we observe that for particles with a higher natural frequency $\Omega_{\rm p}$ (namely lighter and smaller particles), the value of $U_0$ and of $P$ in Eq. (\ref{Eq:KP}) decrease. As a consequence, the bands of allowed frequencies are broadened. Ideally, in the limiting case of particles with zero mass, the natural frequency is infinite and $P$ and $U_0$ vanish. Therefore equation (\ref{Eq:KP}) gets simplified dramatically and becomes $\cos(kd) = \cos(\alpha_\omega d)$. This implies
\begin{equation}
\omega(k)=\Omega_\mathrm{v}\left(k+\frac{2n\pi}{d}\right),\quad n\in\mathbb{Z}
\label{Eq:trackingKW}
\end{equation}
that is just vortex wave dispersion relation, but repeated with period $k_d=2\pi/d$. In other words, light and small particles can follow the filament without modifying the vortex waves. On the contrary, particle inertia reduces the excited frequencies (in absolute value) of vortex excitations. This fact (actually coming from simple linear physics) should be taken into account, when the Kelvin waves are tried to be measured experimentally.

In this work we did not take into account the relevance of buoyancy effects for light and heavy particles. We can estimate it by comparing the buoyancy force  $F_\mathrm{b} = (M_\mathrm{p} - M_0)\tilde{g}$, where $\tilde{g}\sim 9.8m/s^2$ is the gravitational acceleration, with the Magnus force that drives the particles $F_\mathrm{M}=\frac{3}{2}\rho\Gamma a_\mathrm{p}u$, where $u$ is the typical particle velocity estimated as $u \sim\Omega_\mathrm{p} a_\mathrm{p}$. It turns out that $F_\mathrm{b}/F_\mathrm{M}=C\left( \mathcal{M}-1 \right)\left( 2\mathcal{M} +1 \right)$, where $C=\frac{32}{81}\pi^2\frac{\tilde{g}a_\mathrm{p}^3}{\Gamma^2}$. This expression strongly depends on the particle size. For instance, given that the quantum of circulation in superfluid helium is $\Gamma\sim10^{-7}m^2/s$, we get that $C\sim 4\cdot10^{-3}$ for a particle of size $a_\mathrm{p}=1\mu m$ and therefore the buoyancy is negligible. However $C$ becomes of order $1$ for a particle of size $a_\mathrm{p}=7 \mu m$. We conclude that small and light particles would be the most suitable for tracking the vortex excitations.

Several questions can be immediately raised. If particles are not actually equally distributed along the vortex but instead they present some randomness, vortex waves will then propagate in a disordered media. It will be natural then to study the possibility of Anderson localization in such a system \cite{AndersonLocOriginal,Kramer_1993Localization}. Such situation could perhaps appear if the vortex lines are excited by external means, for instance close the onset of the Donnelly-Glaberson instability \cite{ChenDonellyInstability,Glaberson}. 

The physical system studied in this work is a first idealized picture of what happens in real superfluid helium experiments. The most evident difference is that the size of particles is typically orders of magnitude larger than the vortex core size ($\Rp\sim 10^4\xi$). However, the prediction (\ref{Eq:KP}) comes from an asymptotic theory in which $ka_\mathrm{p}\ll1$ and particles can be considered point-like, independently of the functional form of $\Omega_\mathrm{v}(k)$. Therefore, we expect that our result should still apply for wave lengths larger than the particle size.
Such long waves are indeed observed in experiments \cite{FondaKWExp}. In particular, the fact that particle inertia does not affect the (low) frequency Kelvin waves should be still valid. 
A more quantitative prediction for vortices in He II would be always Eq. (\ref{Eq:KP}), but with $\alpha_\omega$ such that $\omega = \Omega_\mathrm{He}(\alpha_\omega)$, where $\Omega_\mathrm{He}(k)$ is the true vortex excitations dispersion relation in superfluid helium. In any case all the main conclusions remain valid, since the analogy with a crystal is independent of $\Omega_\mathrm{v}(k)$. Moreover, the behaviour at large scales is expected to work quantitatively also for superfluid helium vortices because $\Omega_\mathrm{He}(ka_0\rightarrow 0)\sim\Omega_\mathrm{KW}(ka_0)$.

Furthermore, we have used arrays of particles with all identical masses. Instead, in actual experiments there is not a perfect control on the mass and size of particles. In particular, the mass distribution of particles could be poly-dispersed. In this case, new gaps in the dispersion relation are opened revealing much more complex configurations. 
A preliminary numerical study confirms this behavior and it will be reported in a future work. In any case, the basic interaction between one particle and vortex waves remains the same regardless the presence of some disorder. Therefore, large-scale Kelvin waves are not disturbed by the particles. Studying in detail the effects of different species of particles trapped in a vortex can be done systematically in the same spirit of the effective theory developed in the present work, for example adapting tight-binding models \cite{kittel1976introduction} to the vortex-particles system. We think that this is a worthy research direction that could establish new and deeper connections with concepts already known in solid state physics, introducing a plethora of novel phenomena in the framework of quantum fluids. 

Last but not least, note that the basic equations considered in this work to build up the effective model are based on classical hydrodynamics. Therefore, one could expect that most of the phenomenology remains valid in a classical fluid provided that a mechanism to sustain a vortex exists. Such mechanism could be for instance provided by two co-rotating propellers at moderate speeds.
Since these systems are achievable in much less extreme conditions than in cold superfluid helium and because the manipulation of particle parameters is much simpler, it could be possible to build analogs of solid state physics phenomena by using classical fluid experiments.

\appendix
\section{Numerical scheme and parameters}
Equations (\ref{Eq:GPEParticles}-\ref{Eq:GP}) are solved with a standard pseudo-spectral code and a $4^\mathrm{th}$ order Runge-Kutta scheme for the time stepping in a  3D periodic domain of dimensions $L_\perp\times L_\perp\times L_\parallel$ with $N_\perp\times N_\perp\times N_\parallel$ collocation points.  We set $c=\rho_\infty=1$. 

The ground states with particles and straight vortices are prepared separately by performing imaginary time evolution of the GP equation. In order to have an initial state with zero global circulation (and therefore ensure periodic boundary conditions) we need to add in the computational box three image vortices with alternating charges. The state with bundles of $N_\mathrm{v}=4$ vortices (Fig.\ref{Fig:3Dsketch}.4) is prepared imposing a phase jump of $2N_\mathrm{v}\pi$ around a vortex (including its images). Then, imaginary time evolution of GP equation is performed for a time $\sim 150\tau$, so that the vortex filaments separate and the bundles form. KWs are generated from the state with straight vortices slightly shifting each $xy$ plane of the computational domain. Then the states with KWs and particles are multiplied to obtain the wished initial condition. Just one vortex filament is loaded with particles, while the three other images remain bare. The initial condition is evolved for a short time ($\sim40\tau$) using GP without the particle dynamics in order to adapt the system. 

The particle potential is a smoothed hat-function $\Vp(r)=\frac{V_0}{2}(1-\tanh\left[\frac{r^2 -\eta^2}{4\Delta l^2}\right])$ and the mass displaced by the particle is measured as $M_\mathrm{w}^0=\rho_\infty L_\perp L_\parallel^2(1-\int |\psi_\mathrm{p}|^2\,\mathrm{d}\mathbf{x}/\int |\psi_\infty|^2\,\mathrm{d}\mathbf{x})$, where $\psi_\mathrm{p}$ is the steady state with just one particle. Since the particle boundaries are not sharp, we measure the particle radius as $\Rp=(3M_\mathrm{p}^0/4\pi\rho_\infty)^\frac{1}{3}$ for given values of the numerical parameters $\eta$ and $\Delta l$. For all the particles $V_0=20$. The parameters used are the following. 
For $\Rp=2.7\xi$:  $\eta=\xi$ and $\Delta l=0.75\xi$.
For $\Rp=7.6\xi$:  $\eta=2\xi$ and $\Delta l=2.5\xi$. 
For $\Rp=13.1\xi$:  $\eta=10\xi$ and  $\Delta l=2.8\xi$. 

The parameter $r_0$ of the potential $V_\mathrm{rep}^{ij}=\varepsilon(r_0/|\mathbf{q}_i-\mathbf{q}_j|^{12})$ is the radius of the repulsion between particles. The parameter $\varepsilon$ is fixed numerically in order to impose an exact balance between the repulsive force and the GP force $- \int  \Vp(| \x -{\bf q}_i|) \nabla|\psi|^2\, \mathrm{d} \x$ in the ground state with two particles placed at distance $2a_\mathrm{p}$ when $r_0=2a_\mathrm{p}$. The parameters used for the repulsion are the following. For the wires in Fig.\ref{Fig:SpectraCyl}: $r_0=2L_\mathrm{w}/(N_\mathrm{p}a_\mathrm{p})$ and $\varepsilon=4.4 \cdot 10^{-5}$. For the array of particles in Fig.\ref{Fig:KP}: $r_0=4a_\mathrm{p}$ and $\varepsilon=4.4 \cdot 10^{-5}$. For the array of particles in Fig.\ref{Fig:KPbig}: $r_0=2a_\mathrm{p}$ and $\varepsilon=1.7 \cdot 10^{-3}$.

\section{Spatio-temporal spectra}
We use the particle positions to define the spatio-temporal spectra of vortex excitations by computing 
\begin{equation}
S_q(k,\omega)=C_q\left|\int \sum_{j=1}^{N_{\rm p}} q(z_j,t)e^{-i(k z_j+\omega t)}\mathrm{d}t\right|^2,
\label{Eq:SpatioTemporal}
\end{equation}
where $z_j$ the $z$ component of the particle $j$. Similarly, the spatio-temporal spectrum of the superfluid wave function is defined as
\begin{eqnarray}
S_\psi(k,\omega)=C_\psi\left|\int \psi(x,y,z,t)e^{-i(kz+\omega t)}\,\mathrm{d}x\,\mathrm{d}y\,\mathrm{d}z\,\mathrm{d}t\right|^2. \nonumber \\
\label{Eq:SpatioTemporalPsi}
\end{eqnarray}
Note that in Eq. (\ref{Eq:SpatioTemporalPsi}) an average of $\psi$ in the $x$ and $y$ directions is implicit. The normalization constants $C_q$ and $C_\psi$ are set such that the full $(k,\omega)$ integrals of the spatio-temporal spectra is one.
In order to enhance the small scale excitations, in the density plots shown in the present work, both the spectra (\ref{Eq:SpatioTemporal}) and (\ref{Eq:SpatioTemporalPsi}) are further normalized with the frequency-averaged spectra, respectively $\int S_q(k,\omega)\,\mathrm{d}\omega$ and $\int S_\psi(k,\omega)\,\mathrm{d}\omega$. All the color maps shown in the present work are in $\log$-scale.

\section{Derivation of the ``Kronig-Penney" dispersion relation for vortex waves}
We look for a linear wave solution $q(z,t)=\Phi(z)e^{i\omega t}$ of the system (\ref{Eq:system}) and in particular we want to know which frequencies $\omega$ are excited. The function $\Phi(z)$ must satisfy the system 
\begin{eqnarray}
    \frac{\partial^2}{\partial z^2}\Phi(z) + \alpha^2_\omega\Phi(z)=0	\quad (\mathrm{I}) \nonumber \\
    \frac{\partial^2}{\partial z^2}\Phi(z) + \beta^2_\omega\Phi(z)=0	\quad (\mathrm{II})
    \label{Eq:system_phi}
\end{eqnarray}
where $\alpha_\omega$ and $\beta_\omega$ are such that 
\begin{equation}
\Omega_\mathrm{LIA}(\alpha_\omega)=\omega ,\qquad \Omega_\mathrm{LIA}(\beta_\omega)=\omega-\frac{\omega^2}{\Omega_\mathrm{w}},
\label{Eq:alphabeta_implicit}
\end{equation}
that means
\begin{equation}
\alpha_\omega =\sqrt{-\frac{4\pi\omega}{\Gamma\Lambda}}, \qquad
\beta_\omega =\sqrt{\frac{4\pi}{\Gamma\Lambda}\left(\frac{\omega^2}{\Omega_\mathrm{w}}-\omega\right)}
\label{Eq:alphabeta}
\end{equation}

Since the system (\ref{Eq:system_phi}) is a linear and homogeneous differential equation with periodic coefficients of period $d$, it admits a solution in the form $\Phi(z)=e^{ikz}u(z)$, where $u(z)$ is a periodic function of period $d$. The solutions of (\ref{Eq:system_phi}) in the two regions $(\mathrm{I})$ and $(\mathrm{II})$ are 
\begin{eqnarray}
\Phi_\mathrm{I}(z)=e^{ikz}u_\mathrm{I}(z) =e^{ikz}\left[ Ae^{i(\alpha_\omega-k)z} + Be^{-i(\alpha_\omega+k)z}\right] \nonumber \\
\Phi_\mathrm{II}(z)=e^{ikz}u_\mathrm{II}(z) =e^{ikz}\left[  Ce^{i(\beta_\omega-k)z} + De^{-i(\beta_\omega+k)z}\right] \nonumber \\
\label{Eq:sol}
\end{eqnarray}
The coefficients $A$,$B$,$C$,$D$ are fixed by imposing continuity and smoothness of the function $\Phi(z)$ and periodicity of $u(z)$ and its derivative: 
\begin{equation}
\begin{cases}
\Phi_\mathrm{I}(0)=\Phi_\mathrm{II}(0) \\
 \Phi'_\mathrm{I}(0)=\Phi'_\mathrm{II}(0) \\
 u_\mathrm{I}(d-L_\mathrm{w})=u_\mathrm{II}(-L_\mathrm{w}) \\ 
 u'_\mathrm{I}(d-L_\mathrm{w})=u'_\mathrm{II}(-L_\mathrm{w}) 
 \end{cases}
\label{Eq:condition}
\end{equation}
The system (\ref{Eq:condition}) is a homogeneous linear system for the variables $A$,$B$,$C$,$D$. It admits non-trivial solutions only if the determinant of the coefficients is equal to zero. This imply the following condition
\begin{eqnarray}
\cos(kd) &=&\cos(\beta_\omega L_\mathrm{w})\cos(\alpha_\omega(d-L_\mathrm{w})) \nonumber \\ 
 &&-\frac{\alpha_\omega^2+\beta_\omega^2}{2\alpha_\omega\beta_\omega}\sin(\beta_\omega L_\mathrm{w})\sin(\alpha_\omega(d-L_\mathrm{w})) \nonumber \\
\label{Eq:FullKP}
\end{eqnarray}
which determines implicitly the dispersion relation $\omega(k)$. Such expression is structurally identical to the standard Kronig-Penney condition but the function $\alpha_\omega$ and $\beta_\omega$ are different. The limit $L_\mathrm{w}\rightarrow 0$ is applied to Eq. (\ref{Eq:FullKP}), substituting at the same time the mass of the massive vortex segment $M_\mathrm{w}^\mathrm{eff}$ with the mass of the particle $M_\mathrm{p}^\mathrm{eff}$. In this way the system becomes a vortex filament loaded with massive point-particles (see Fig.\ref{Fig:Sketch}). The limit implies $\beta_\omega\rightarrow \infty$, $\beta_\omega L_\mathrm{w}\rightarrow 0$, $\sin(\beta_\omega L_\mathrm{w})\sim \beta_\omega L_\mathrm{w}$, $\alpha_\omega \ll \beta_\omega$ and $\beta^2_\omega L_\mathrm{w}\sim 6\pi a_\mathrm{p}\omega^2/\Lambda\Gamma\Omega_\mathrm{p}$, so that Eq. (\ref{Eq:FullKP}) becomes Eq. (\ref{Eq:KP}).

The previous result can be extended to the case of more realistic vortex waves with some caveat. We can formally rewrite the model (\ref{Eq:system}) as
\begin{eqnarray}
    \dot{q}(z,t)=i\hat{\mathcal{L}}_\mathrm{v}[q(z,t)]	\quad  (\mathrm{I}) \nonumber \\
    \ddot{q}(z,t)= i\Omega_\mathrm{w}\left[ \dot{q}(z,t) -  i\hat{\mathcal{L}}_\mathrm{v}[q(z,t)]\right]	\quad (\mathrm{II})
    \label{Eq:systemKW}
\end{eqnarray}
where $\hat{\mathcal{L}}_\mathrm{v}$ is the linear non-local differential operator that generates the vortex wave dispersion relation $\Omega_\mathrm{v}(k)$. Namely, calling $s(z,t)=\sum_k s_k(t)e^{ikz}$ the wave operator simply reads
\begin{equation}
\hat{\mathcal{L}}_\mathrm{v}[s(z,t)]=\sum_k \Omega_\mathrm{v}(k)s_k(t) e^{ikz}.
\label{Eq:LKW}
\end{equation}
The system (\ref{Eq:system_phi}) thus becomes
\begin{eqnarray}
   \hat{\mathcal{L}}_\mathrm{V}[\Phi(z)] -\omega\Phi(z)=0	\quad (\mathrm{I}) \nonumber \\
    \hat{\mathcal{L}}_\mathrm{V}[\Phi(z)] - \left(\omega - \frac{\omega^2}{\Omega_\mathrm{v}} \right)\Phi(z)=0	\quad (\mathrm{II})
    \label{Eq:system_phiKW}
\end{eqnarray}
The functions (\ref{Eq:sol}) are still a solution of (\ref{Eq:system_phiKW}), but now $\alpha_\omega$ and $\beta_\omega$ are defined as
\begin{equation}
 \Omega_\mathrm{v}(\alpha_\omega)=\omega, \qquad  \Omega_\mathrm{v}(\beta_\omega) = \left(\omega-\frac{\omega^2}{\Omega_\mathrm{w}}\right).
\label{Eq:alphabetaKW}
\end{equation}
In general such equations cannot be inverted explicitly, but $\alpha_\omega$ and $\beta_\omega$ can be found numerically. In particular the inversion is intended with respect to $\Omega_\mathrm{v}(k>0)$. The functions $\alpha_\omega$ and $\beta_\omega$ are well defined (at least for $\omega/\Gamma>0$) because any model for the self-induced velocity of a vortex generates a dispersion relation $\Omega_\mathrm{v}(k)$ that is monotonically increasing for positive $k$. For evaluating the limit $L_\mathrm{w}\rightarrow 0$, $M_\mathrm{w}^\mathrm{eff}\rightarrow M_\mathrm{p}^\mathrm{eff}$, we note that $\lim_{L_\mathrm{w}\rightarrow 0} \Omega_\mathrm{v}(\beta_\omega)=\infty$. Therefore, we can explicitly use the asymptotics of $\Omega_\mathrm{v}(k)$ for large $k$, that is just free particle dispersion relation (\ref{Eq:freeParticles}) and can be inverted explicitly: 
\begin{equation}
\beta_\omega \underset{L_\mathrm{w}\rightarrow 0}{\longrightarrow} \sqrt{\frac{4\pi\omega^2}{\Gamma\Omega_\mathrm{w}}},
\label{Eq:LIAKW}
\end{equation}
so that $\beta^2_\omega L_\mathrm{w}\sim 6\pi a_\mathrm{p}/\Gamma\Omega_\mathrm{p}$. In this way we recover Eq. (\ref{Eq:KP}), with $\alpha_\omega$ defined as in (\ref{Eq:alphabetaKW}) and the amplification factor $P$ is now independent of any free parameter:
\begin{equation}
P=\frac{3\pi d a_\mathrm{p}}{\Gamma\Omega_\mathrm{p}}.
\label{Eq:NewP}
\end{equation}

\begin{acknowledgments}  
We acknowledge useful scientific discussions with Davide Proment and Vishwanath Shukla. U.G. and G.K were supported by Agence Nationale de la Recherche through the project GIANTE ANR-18-CE30-0020-01. Computations were carried out on the M\'esocentre SIGAMM hosted at the Observatoire de la C\^ote d'Azur and the French HPC Cluster OCCIGEN through the GENCI allocation A0042A10385. SN is supported by the Chaire d'Excellence IDEX, UCA. SN and GK are also supported by the EU Horizon 2020 Marie Curie project HALT.
\end{acknowledgments}

%\bibliographystyle{apsrev4-1}
%\bibliography{TrackingVorticesWithParticles.bib}

%merlin.mbs apsrev4-1.bst 2010-07-25 4.21a (PWD, AO, DPC) hacked
%Control: key (0)
%Control: author (72) initials jnrlst
%Control: editor formatted (1) identically to author
%Control: production of article title (-1) disabled
%Control: page (0) single
%Control: year (1) truncated
%Control: production of eprint (0) enabled
%

\end{document}